\documentclass{article}

\usepackage{pdfpages} 
\usepackage{pgffor} 

\makeatletter
\AtBeginDocument{\let\LS@rot\@undefined}
\makeatother

\def\supplementfilename{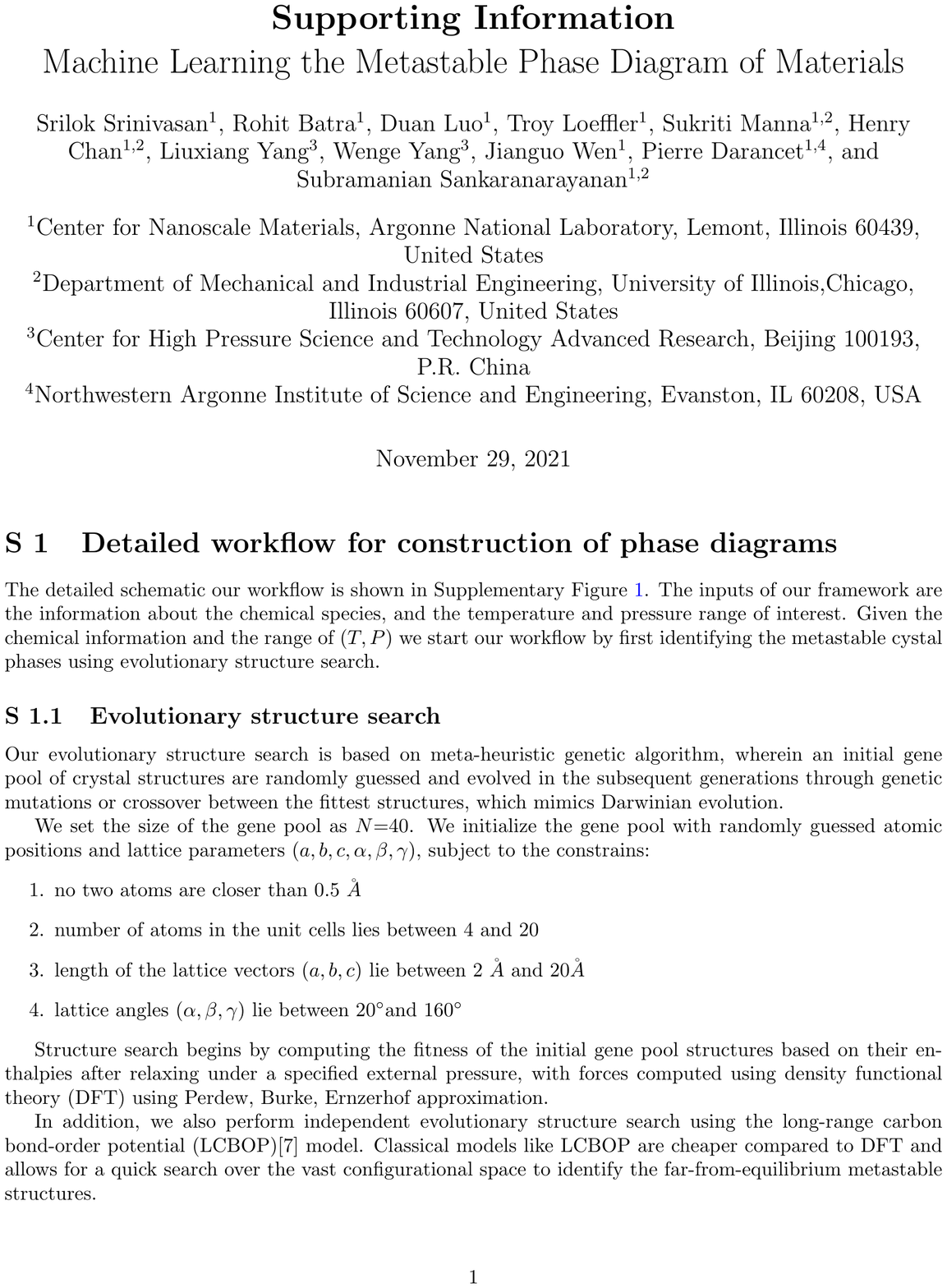}

\pdfximage{\supplementfilename}
\def\numbersupplementpages{\the\pdflastximagepages}

\newif\ifarXiv
\arXivtrue 

\usepackage[utf8]{inputenc}
\usepackage{geometry}
\usepackage{graphicx}
\usepackage{xcolor}
\usepackage{bm}
\usepackage{amsmath}
\usepackage[normalem]{ulem}
\usepackage{graphicx}
\graphicspath{ {images/} }
\usepackage{color,soul}
\usepackage{gensymb}
\usepackage[colorlinks=true, citecolor=black, urlcolor=blue,linkcolor=blue]{hyperref}
\usepackage[square,numbers]{natbib}
\usepackage{authblk}
\usepackage{mathtools}

\begin{document}

\title{\textbf{Machine Learning the Metastable Phase Diagram of Materials}}
\author[1]{Srilok Srinivasan}
\author[1]{Rohit Batra}
\author[1]{Duan Luo}
\author[1]{Troy Loeffler}
\author[1,2]{Sukriti Manna}
\author[1,2]{Henry Chan}
\author[3]{Liuxiang Yang} 
\author[3]{Wenge Yang}
\author[1]{Jianguo Wen}
\author[1,4]{Pierre Darancet}
\author[1,2]{Subramanian Sankaranarayanan}
\affil[1]{Center for Nanoscale Materials, Argonne National Laboratory, Lemont, Illinois 60439, United States}
\affil[2]{Department of Mechanical and Industrial Engineering, University of Illinois,Chicago, Illinois 60607, United States}
\affil[3]{Center for High Pressure Science and Technology Advanced Research, Beijing 100193, P.R. China}
\affil[4]{Northwestern Argonne Institute of Science and Engineering, Evanston, IL 60208, USA}

\date{\today}

\maketitle
\begin{abstract}
A central feature of materials synthesis is the concept of phase diagrams. Phase diagrams are an invaluable tool for material synthesis and provide information on the phases of the material at any given thermodynamic condition (i.e., state variables such as pressure, temperature and composition). Conventional phase diagram generation involves experimentation to provide an initial estimate of the set of thermodynamically accessible phases and their boundaries, followed by use of phenomenological models to interpolate between the available experimental data points and extrapolate to experimentally inaccessible regions. Such an approach, combined with high throughput first-principles calculations and data-mining techniques, has led to exhaustive thermodynamic databases (e.g. compatible with the CALPHAD method), albeit focused on the reduced set of phases observed at distinct thermodynamic equilibria. In contrast, materials during their synthesis, operation, or processing, may not reach their thermodynamic equilibrium state but, instead, remain trapped in a local (metastable) free energy minimum, that may exhibit desirable properties. Mapping these metastable phases and their thermodynamic behavior is highly desirable but currently lacking, due to the vast configurational landscape. Here, we introduce an automated workflow that integrates first principles physics and atomistic simulations with machine learning (ML), and high-performance computing to allow rapid exploration of the metastable phases of a given elemental composition. Using a representative material, carbon, with a vast number of metastable phases without parent in equilibrium, we demonstrate automatic mapping of hundreds of metastable states ranging from near equilibrium to those far-from-equilibrium (500 meV/atom). Moreover, we incorporate the free energy calculations into a neural-network-based learning of the equations of state that allows for construction of metastable phase diagrams. High temperature high pressure experiments using a diamond anvil cell on graphite sample coupled with high-resolution transmission electron microscopy (HRTEM) are used to validate our metastable phase predictions.  Our introduced approach is general and broadly applicable to single and multi-component systems.
\end{abstract}

\maketitle

\section{Introduction}

Materials synthesis has traditionally relied on “thermodynamic phase diagrams” to provide information about the stable phases as a function of various intensive state properties such as temperature, pressure, and chemical composition. The conventional method for generating a phase diagram involves experimentation to provide an initial estimate of phase boundaries followed by the use of phenomenological models to interpolate the available experimental data points and extrapolate to experimentally inaccessible regions. Such an approach combined with atomistic simulations and recent data-mining techniques has led to well-established exhaustive thermodynamic databases~\cite{DatabasePhaseDiagramANDERSSON,OpenCalphad,ThermodynamicDatabaseAxelVandeWalle} for different materials—albeit limited to phases observed near thermodynamic equilibria. However, following material synthesis and processing, or during operation, most materials may be trapped in local minima of the energy landscape, that is, in metastable states (see Figure \ref{fig:Schmatic}(a)). Solid carbon is a prototypical system exhibiting such behavior, with large number of known metastable allotropes at room temperature and atmospheric pressure. Importantly, these allotropes have contrasting properties ranging from metals~\cite{MetallicAllotropesPRL,HoffmanHypoStruct,2DGrapheneToGraphyne,K6}, semiconductors~\cite{PhenylRingSciRep}, topological insulators~\cite{CrommieLouieGNRtopo,StevenLouie-TopoGNRPRL,StevenLouieTopoNanoLetter}, and wide band gap insulators~\cite{DiamondAsElectronicMaterial}. There is likely a vast and rich phase space of metastable structures for multi-component materials, with some exhibiting exotic and potentially desirable properties. The demand for such materials motivates the move beyond the traditionally explored area of \textit{near-equilibrium} materials.  Towards this goal, the generation of exhaustive datasets of ``metastable phase diagrams'', mapping the equation of states for phases without parent in thermodynamic equilibrium, is highly desirable but has remained elusive.

Creating a phase map for metastable materials is a non-trivial and data-intensive task. The first challenge is to employ an efficient structure optimization algorithm capable of identifying both global (ground state) and local (metastable) minima  of the energy landscapes in the configurational space. 
The next challenge is to map the free energy surface (i.e. the equation of state) for each of these metastable phases as a function of the intensive thermodynamic state variables ($P$, $T$ and $X$), over the range in which the phase information is desired. As this information is usually discretized into a finite grid of state variables, one needs to find the free energies($G(T,P,X)$) of individual metastable phases at each grid point. This step quickly becomes computationally prohibitive for large numbers of metastable configurations, and, in practice, requires a surrogate model, to approximate the free energy calculations of more expensive first-principles based approach (e.g. ab-initio molecular dynamics). The final, major challenge after the equation of state for all the phases are computed is to classify and identify the phase boundaries at varying degrees of non-equilibrium i.e. the areas of the phase diagram in which a metastable structure is dynamically decoupled from lower energy structures.

Here, we report an automated framework that addresses the above challenges by integrating a genetic algorithm with first-principles calculations, classical molecular dynamics simulations, machine learning (ML), and high-performance computing to allow the generation and exploration of the metastable materials. Our framework allows the automatic discovery, identification, and exploration of the metastable phases of a material, and ‘learns’ their equations of state through a deep neural network. To test the efficacy of our framework, we use the representative and highly-significant example of carbon --a system well-known to exhibit a large number of metastable allotropes –and map its metastable phase diagram in a large range of temperatures (0-3000 K), pressure (0- 100 GPa) and excess free energy (up to 500 meV/atom above thermodynamic equilibrium). Importantly, we show that the proximal phases to thermodynamic equilibrium (within 50 meV/atom) can be observed experimentally in high pressure high temperature (HPHT) processing of  graphite. In particular, we identify a new cubic-diaphite metastable configuration that explains the diffraction pattern of the previously reported \textit{n}-diamond~\cite{RN1}, demonstrating the potential of our approach to guide the synthesis of materials beyond equilibrium. 

\section{Method}

\begin{figure}
\noindent\makebox[\textwidth]{\includegraphics[width=0.90\textwidth]{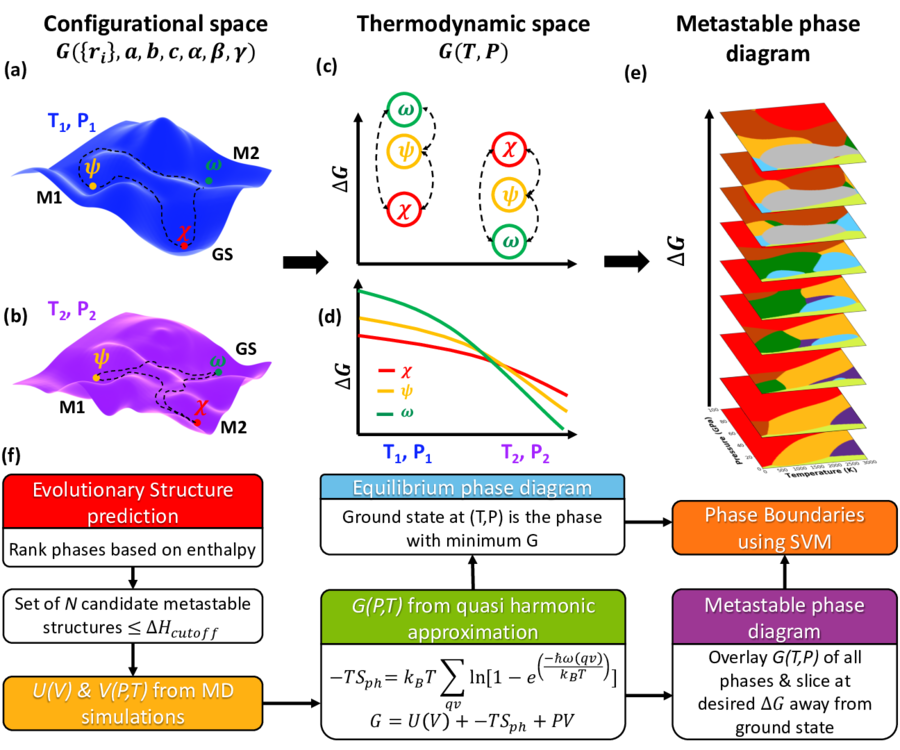}}
\caption{(a) and (b) Schematic illustration of the free energy landscape in the configurational space at different conditions $(T_1,P_1)$, $(T_2,P_2)$. The phases corresponding to the minima are labeled $\chi, \psi, \omega$. GS, M1 and M2 stand for ground state, near-equilibrium and far-from equilibrium metastable phases; (c) Graph representation of the energy landscape. Nodes correspond to the phases and the edges contain the barrier height; (d) Equation of state for $\chi, \psi$ and $\omega$ (e) Illustration of the metastable phase diagram as a function of $\Delta G$; (f) Our workflow to identify metastable configurations and construct the metastable phase diagram}
\label{fig:Schmatic}
\end{figure}

Our workflow is summarized in Figure \ref{fig:Schmatic}. We construct  metastable phase diagrams with the chemical information of the periodic system as input, along with the range of pressure and temperature of interest. 

The ground and metastable states at a given set of thermodynamic conditions $(T,P)$ correspond to global and local minima of the free energy in the configurational space, $G(\{r_i\},a,b,c,\alpha,\beta,\gamma)$, where $a,b,c,\alpha,\beta,\gamma$ are the lattice parameters and $\{r_i\}$ are the position of the basis atoms. As explained in details below, we identify the metastable phases by sampling the energy landscapes at fixed thermodynamic conditions. We then compute, at the identified minima, the Gibbs free energy in the thermodynamic space as function of intensive variables $G(T,P)$; the free energy and the relative energetic ordering of its minima varies with $(T,P)$ as illustrated in Figure \ref{fig:Schmatic} (a),(b). Upon identification, the free energy and stability of these phases at $(T,P)$ is  represented as a graph (Figure \ref{fig:Schmatic}(c)) with nodes corresponding to the free energy of the phases and the edges to the free energy barrier connecting them. This discrete thermodynamics representation is made continuous as a function of $(T,P)$, and the crossing points in equation of states automatically identified. Finally, we generate the full metastable phase diagram, $\mathcal{P}\left(T,P,\Delta G\right)$, where  $\mathcal{P}$ is the most energetic phase within a free energy $\Delta G$ of the ground state at a given $\left(T,P,\Delta G\right)$.

\subsection{Evolutionary structure prediction}
The first step in our workflow is to identify the periodic structures that are energetically favorable for a given chemical composition. We use an evolutionary search based on genetic algorithm -- known to be efficient for periodic systems~\cite{USPEXWhyEvol.Algo.Works.Oganov,NaoMaromGAtor,GASPHenningRevard2014}. Briefly, evolutionary algorithms aspire to optimize the atomic arrangement {$r_1$,$r_2$,...$r_n$} and the lattice parameters ($a,b,c,\alpha,\beta,\gamma$) of a \textit{population} of structures over different regions of the energy landscape through genetic variations and selections over successive iterations. The structure corresponding to the global minima is the ground state equilibrium structure. Conversely, the metastable phases at a finite temperature and pressure correspond to the local minima of the Gibbs free energy landscape $G(\{r_i\},a,b,c,\alpha,\beta,\gamma)$. Hence, evolutionary algorithms are naturally suited to locate candidate metastable phases over the configurational space by evolving a pool of structures at the same time. Although $G$ includes both the temperature(-$TS$) and pressure ($PV$) contributions, for computational cost efficiency, we only include the effect of finite pressure in the selection of the offspring structures, by optimizing enthalpy at 0 K and fixed pressure, $H(T=0\,K,P)$. The entropic contribution will be computed in the subsequent steps of our workflow.

We perform evolutionary structure search at several different pressures ($P=0\,GPa, P=10\,GPa$ \&\ $P=100\,GPa$) independently by minimizing $H(T=0\,K,P)$. The search is initiated with a population of $N=40$ randomly generated crystal structure satisfying the following geometric constrains: (i) no two atoms are closer than 0.5 \AA, (ii) total number of atoms in the unit cells lies between 4 and 20, (iii) length of the lattice vectors ($a,b,c$) lie between 2 \AA\ and 20 \AA\, (iv) lattice angles ($\alpha,\beta,\gamma$) lie between 20\degree  and 160\degree. Each member of the population is locally relaxed using density functional theory with Perdew, Burke, Ernzerhof approximation before computing its enthalpy. Including such chemically informed local optimization accelerates the search for minima and avoids sampling unphysical configurations. In addition, we also perform independent evolutionary structure searches using the long-range carbon bond-order potential (LCBOP)\cite{RN12} model. Classical models like LCBOP are cheaper compared to DFT and allows for a quick search over the vast configurational space.

During the search, “parent” structures are selected from the current population with probabilities proportional to their fitness, computed based on $H(T=0\,K,P)$ (see supporting information S1). The phase with the lowest(highest) enthalpy has the highest(lowest) fitness. Genetic operations are performed on the parents to produce new “offspring” crystal structures. Subsequently, a new generation is constructed from $N$ best structures from the previous generation and the new offspring structures. The above cycle is repeated until the enthalpy of the top $N/8$ structures are within the desired tolerance. Further details on the structure search algorithm such as the selection criteria, types of genetic operations and parameters used, are provided in the supporting information (section S1). All the new phases encountered during the search and their corresponding enthalpy values are recorded. Candidate phases for free energy calculations are identified from a collated list of structures from several independent evolutionary structure search at different pressures. 

\subsection{Metastable phase identification}

\begin{figure*}
\noindent\makebox[\textwidth]{\includegraphics[width=0.95\textwidth]{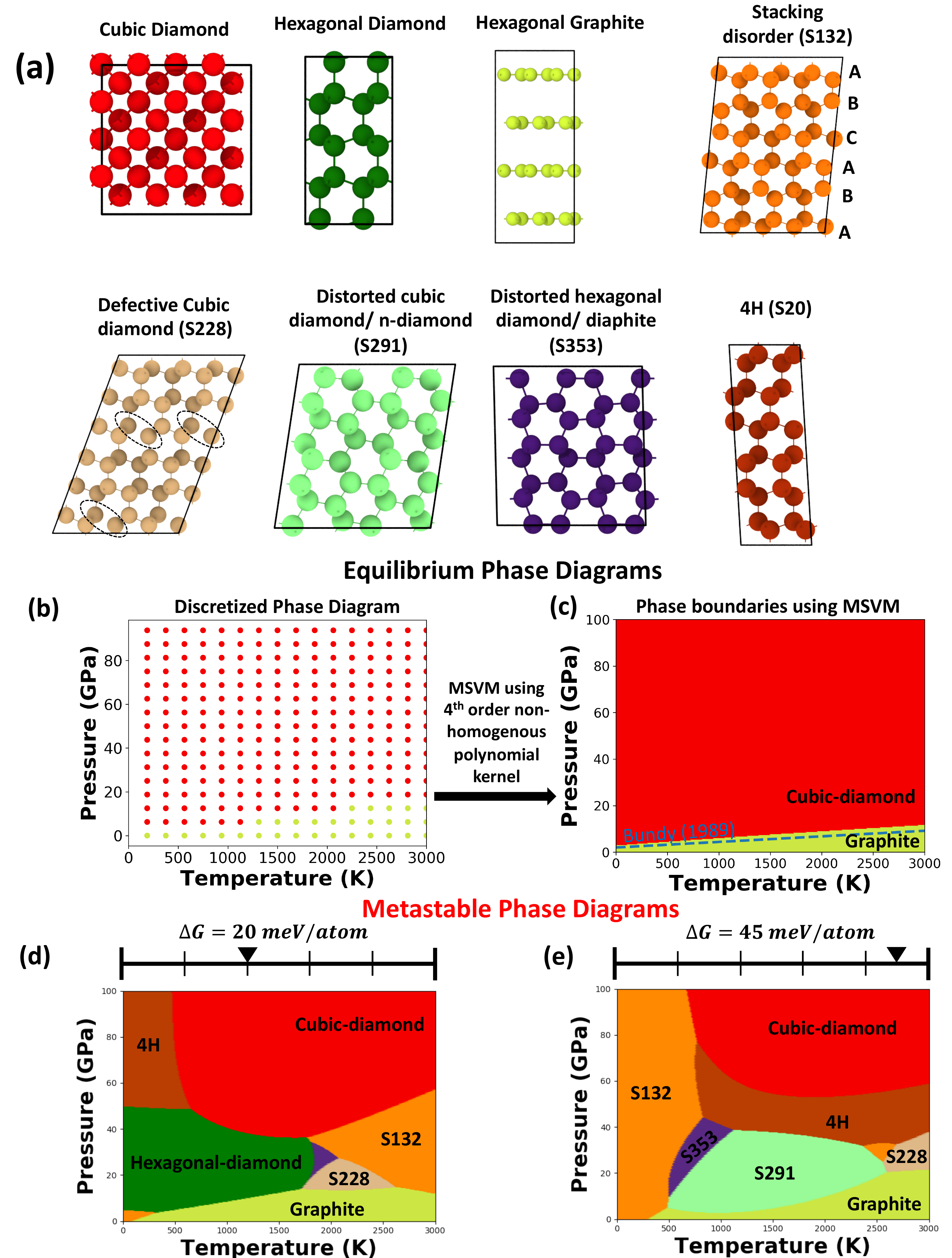}}
\caption{(a) Near equilibrium phases identified by evolutionary structure search; (b) discretized equilibrium phase diagram generated using the above workflow; (c) equilibrium phase diagram with boundary fitted using MSVM. Equilibrium phase diagrams matches with the experimental phase diagram\cite{RN11,RN12}; (d) \& (e) The metastable phase diagram (at a $\Delta G$ of 25 meV and 50 meV respectively) shows the appearance of metastable phases listed in the panel at high temperatures and moderate pressures.}
\label{fig:phasediagram}
\end{figure*}

While any point $(T,P)$ in the equilibrium phase diagram shows the ground state, there exists several local minima (metastable phases) in the configurational space ($\{r_i\},a,b,c,\alpha,\beta,\gamma$) separated by a finite free energy difference $\Delta G$ (Figure \ref{fig:Schmatic}(c),(d)). Figure \ref{fig:phasediagram}(a) depicts some of representative metastable structures found during structure search. Cubic diamond and hexagonal graphite appear in the experimental equilibrium phase diagram~\cite{RN11} of carbon. Apart from the equilibrium phases, our evolutionary structure search algorithm also identified metastable structures like the hexagonal diamond, 4H phase, other stacking combinations of cubic and hexagonal diamond (stacking disorder), distorted cubic diamond, distorted hexagonal diamond (diaphite)
, which are also observed in our HPHT experiments (see below). At a given pressure, the structure with minimum enthalpy ($H_{ground}$) is the ground state at 0 K -- in the case of carbon, graphite at 0 GPa. At 0 K, we have $ H(T=0\,K,P) = G(T=0\,K,P)$. Hence, we can use a cutoff criteria based on $H(T=0\,K,P)$ to screen the candidate metastable phases for the subsequent free energy calculation. We define a $\Delta H_\mathrm{cut-off}$ and neglect the structures whose enthalpy is more than $\Delta H_\mathrm{cut-off}$  from the ground state. Upon convergence of the evolutionary structure search, we only select the structures that satisfy $H< H_\mathrm{ground}$+$\Delta H_\mathrm{cut-off}$ for further analysis. In the present work, we set $\Delta H_\mathrm{cut-off} = 670 \,meV/atom$, comparable to the excess enthalpy of C60 fullerene ($\Delta H_\mathrm{C60} = 608\, meV/atom$)  that, we hypothesize, should be large enough to include the thermodynamically relevant metastable structures. Among the selected structures, we group geometrically similar and layered structures (for example hexagonal graphite, orthorhombic graphite, rhombohedral graphite) based on the radial distribution function, angular distribution function (see supporting information, section S1.2) which further reduces the number of candidate structures for free energy calculation. After performing the above selection and grouping of structures, we narrow down to 505 candidate metastable structures for free energy calculation.

\subsection{Free energy calculation and discrete phase diagram }
The candidate structures obtained after performing the previous steps are entirely based on the enthalpy values at 0 K. However, the metastability of a structure at a finite temperature is determined based on the Gibbs free energy $G(T,P)$. The continuous axis of temperature and pressure in the phase diagram are discretized into 2D grid (in this case 16x16) within the range of interest as shown in Figure \ref{fig:phasediagram}(b) to reduce the number of free energy calculations. The Gibbs free energy of each candidate screened from the previous step is computed at each of the grid points. At a given temperature and pressure:
\begin{equation}\label{1stEq} 
    G(T,P) = H(T,P) - TS(T,P).
\end{equation}

In solids with few atomic components, the vibrational contribution to the entropy is the dominant one~\cite{ReviewModernPhysicsCederVandeWaale}, and hence we make the approximation:
\begin{equation}\label{2ndEq} 
    S(T,P) \approx S_{\textrm{vibrational}}(T,P), 
\end{equation}

If the atomic vibrations are modeled as harmonic oscillators, it follows that 
\begin{equation}\label{3rdEq} 
    -TS_{\textrm{vibrational}}(T,P) = F_{\textrm{Harmonic}} - U_{\textrm{Harmonic}}  = k_BT\sum_{qv}\ln[1-\exp(-\dfrac{\hbar\omega(qv)}{k_BT}]
\end{equation}
The enthalpy of the system, $H(T,P)$, is estimated at each of the grid points from MD simulations by equilibrating under NPT ensemble, while the entropic part of the free energy is computed from the phonon spectra using Eq. (3). The phonon spectra and the vibrational free energies are calculated at the equilibrium density obtained from MD simulations at the corresponding $T$ and $P$. Further details on the MD simulations and the vibrational free energy calculations can be found in supporting information section S1.3. The MD simulations are performed using LAMMPS package~\cite{RN9} and the phonon spectra is calculated using the PHONOPY package\cite{RN10}. 

Once we have the free energies of all the structures, the discretized phase diagram is constructed by comparing the $G(T,P)$ of the candidate structures at each point in the 2D grid. 

\subsection{Phase-dependent equations of state through Deep Neural Networks}

As shown below, recently developed machine learning (ML) methods~\cite{lecun2015deep, ramprasad2017machine, meredig2014combinatorial} for developing inter-atomic potentials~\cite{behler2016perspective, botu2017machine, huan2017universal, pun2019physically} or estimating atomistic or molecular properties~\cite{rupp2015machine, xie2018crystal, chen2019graph} can be used to compute Gibbs free energy as a continuous function of $T,P$. In particular, the equation of state of a phase can be predicted directly given only the 0 K structural information of a phase, allowing us to quickly estimate a $(T,P)$ region wherein a specific phase has low Gibbs free energy, and can be potentially realized in the experiments. 

We use a deep neural network (DNN) that takes as an input the many-body tensor representation~\cite{huo2017unified} (MBTR) of a phase, along with $T$ and $P$ information. The DNN is trained on the Gibbs free energy data of 248 phases out of the 505 carbon phases. Regularization techniques, such as dropout and early stopping, were utilized to avoid overfitting. Some important low energy metastable phases, namely, hexagonal diamond, defective cubic diamond (S228) and 4H (S20), were intentionally left out from the training process and were used to evaluate the DNN performance. More details on the DNN architecture, training, and the MBTR descriptor are provided in the supporting information section S3.

\subsection{Phase boundary classification - Discretized to continuous phase diagram}

The final stage in our workflow is to clearly identify the phase boundaries as a function of $\left(T,P,\Delta G\right)$ separating the different phases. Machine learning algorithms like support vector machines (SVMs)\cite{RN29,RN30} which can draw the decision boundaries between different classes of inputs are well suited to automate the estimation of such phase boundaries. SVMs are binary classifiers by definition and one have to resort to decomposition techniques like “one-vs-all” or “one-vs-rest”\cite{RN31} which involves training many classifiers and taking the weighted value of all the output. While such decomposition techniques have been successfully used in the past, it is computationally demanding to train multiple classifiers when there are a large of number phases. Instead, we use a purely multiclass SVM\cite{RN32,RN33,RN34,RN35,RN36} (MSVM), using a non-homogenous 4th order polynomial kernel, which can classify multiple classes without relying on decomposition techniques. Training only one MSVM classifier reduces the computational time tremendously while maintaining the accuracy of the classifiers. The final equilibrium and metastable phase diagram can be generated with the decision boundaries drawn using MSVM (Figure \ref{fig:phasediagram}).

\section{Results}

\subsection{Equilibrium phase diagram}
We first validate our workflow by constructing equilibrium phase diagram and comparing against the experimental graphite-diamond phase boundary \cite{RN11,RN12}. The discretized equilibrium phase diagram (Figure~\ref{fig:phasediagram}(b)) is constructed by comparing the $G(T,P)$ of diamond and graphite, and plotting the phase with a lower $G$ at each grid point. The color of the points correspond to the color of the structures shown in Figure~\ref{fig:phasediagram}(a). As expected, from the experimental phase diagram~\cite{RN11,RN13}, the cubic diamond phase is dominant at high pressure whereas graphite is more stable in the low-pressure region. Importantly, our predicted diamond-graphite phase boundary matches well with the experimental phase boundary (dashed line in Figure~\ref{fig:phasediagram}(c)). 

We also note that $G(T,P)$ for S132 and S353 are very close to diamond at moderate pressures and slightly lower ($\Delta G/k_BT < 0.3$) than diamond and graphite at high temperatures (see supporting information S2). These phases correspond to the stacking disorder phase (orange in Figure~\ref{fig:phasediagram}(a)) consisting of alternating layers of cubic diamond and hexagonal diamond and a diaphitine like distorted hexagonal diamond (purple in Figure \ref{fig:phasediagram}(a)) with two different bond lengths at 1.47 \AA\ and 1.53 \AA. While both the stacking disorder and the diaphitine-like lonsdaleite phases are widely believed to be metastable~\cite{RN23,RN24,RN25,RN26}, our calculations show that they lie near the experimental phase boundary. Incidentally, our theoretically predicted ranges of stability for stacking disorder also matches with the experimental conditions under which they are observed~\cite{RN16,RN18,RN19,RN27,RN28}. For instance, hexagonal diamond (lonsdaleite) containing varying fraction of cubic diamond~\cite{RN24}, alternatively described as a stacking disorder diamond or faulted and twinned cubic diamond~\cite{RN26} has been experimentally synthesized under static compression~\cite{RN16,RN17,RN18,RN19,RN20}, HPHT-treatment~\cite{RN21,RN22} or shock compression~\cite{RN22,RN23,RN24,RN25}.  
These observations are not surprising considering that, near the phase boundaries, the energetic differences between the experimentally reported metastable phases and the stable (cubic diamond, graphite) phases is only $0.3\times k_BT$ or less (see supporting information). Such a small difference increases the likelihood (discussed below) of forming these phases at high temperatures.

\subsection{Metastable phase diagram}

We next construct the metastable phase diagram of carbon. While the phases represented in the equilibrium phase diagram exhibit minimum free energy at a given pressure and temperature,  metastable phases 
are located in valleys of the high dimensional free energy landscape with respect to the structural parameters (refer schematic Figure \ref{fig:Schmatic}(a),(b)). We define the quantity $\Delta G_{GS_i}^{MS_j} = G_{MS_j} - G_{GS_i}$ as the difference in Gibbs free energy between the metastable structure $MS_j$ and the ground state $GS_i $ at given temperature and pressure, with $\Delta G_{GS_i}^{GS_i}(T,P)=0$ and $\Delta G_{GS_i}^{MS_j}(T,P) > 0 $ if $MS_j$ and $GS_i$ are distinct phases. 

The probability of realizing a metastable phase at a given temperature and pressure is directly proportional to $\exp(-\dfrac{\Delta G_{GS_i}^{MS_j}}{k_BT})$. We therefore construct a $\Delta G(T,P)$ surface, the projections of which can be used to derive the metastable phase diagram as a function of the degree of non-equilibrium from the corresponding equilibrium phase. We thus define the metastable phase diagram as the phase diagram obtained by projecting on the $T-P$ plane, the phase with closest $\Delta G_{GS_i}^{MS_j}(T,P)$  value compared to a given $\Delta G$, which is also the measure of degree of non-equilibrium, and satisfies $\Delta G_{GS_i}^{MS_j}(T,P) < \Delta G$.  In other words, by varying $\Delta G$, we are effectively taking slices of the overlaid free energy landscape (Figure \ref{fig:Schmatic}(e) \& Figure \ref{fig:domains}(a)) of all the structures. Experimentally, such phases can be accessed by using pulsed laser heating, in which the system undergoes phase transformation with the pulse providing the energy to transition between local minima of the free energy (Figure \ref{fig:Schmatic}(c)). 

Figure~\ref{fig:phasediagram}(d) and~\ref{fig:phasediagram}(e) shows the metastable phase diagram of carbon at $\Delta G$ equal to 20 meV/atom and 45 meV/atom respectively. At a non-zero $\Delta G$, we see the appearance of new metastable phases in the phase diagram. At much higher values of $\Delta G$ more than 20 metastable phases appear in the metastable phase diagram (section S4 in supporting information). Representing the metastability of the different phases in such a phase diagram offers a wealth of information. One can deduce the temperature-pressure ranges at which a phase is likely to be stabilized and an estimate of minimum excitation energies (from $\Delta G$) required to synthesize a metastable phase, thus offering a systematic approach in designing experiments at favorable conditions for synthesis. 
It is interesting to note that, the neighboring phases in the metastable phase diagram are structurally similar suggesting, and there may exist, a low energy transition pathway connecting them. For example, the diaphite phase, is adjacent to the regular hexagonal diamond structure in the phase diagram at a $\Delta G$= 20 meV/atom.  




\subsection{High Temperature High Pressure Experiments}

\begin{figure*}[!htb]
\noindent\makebox[\textwidth]{\includegraphics[width=\textwidth]{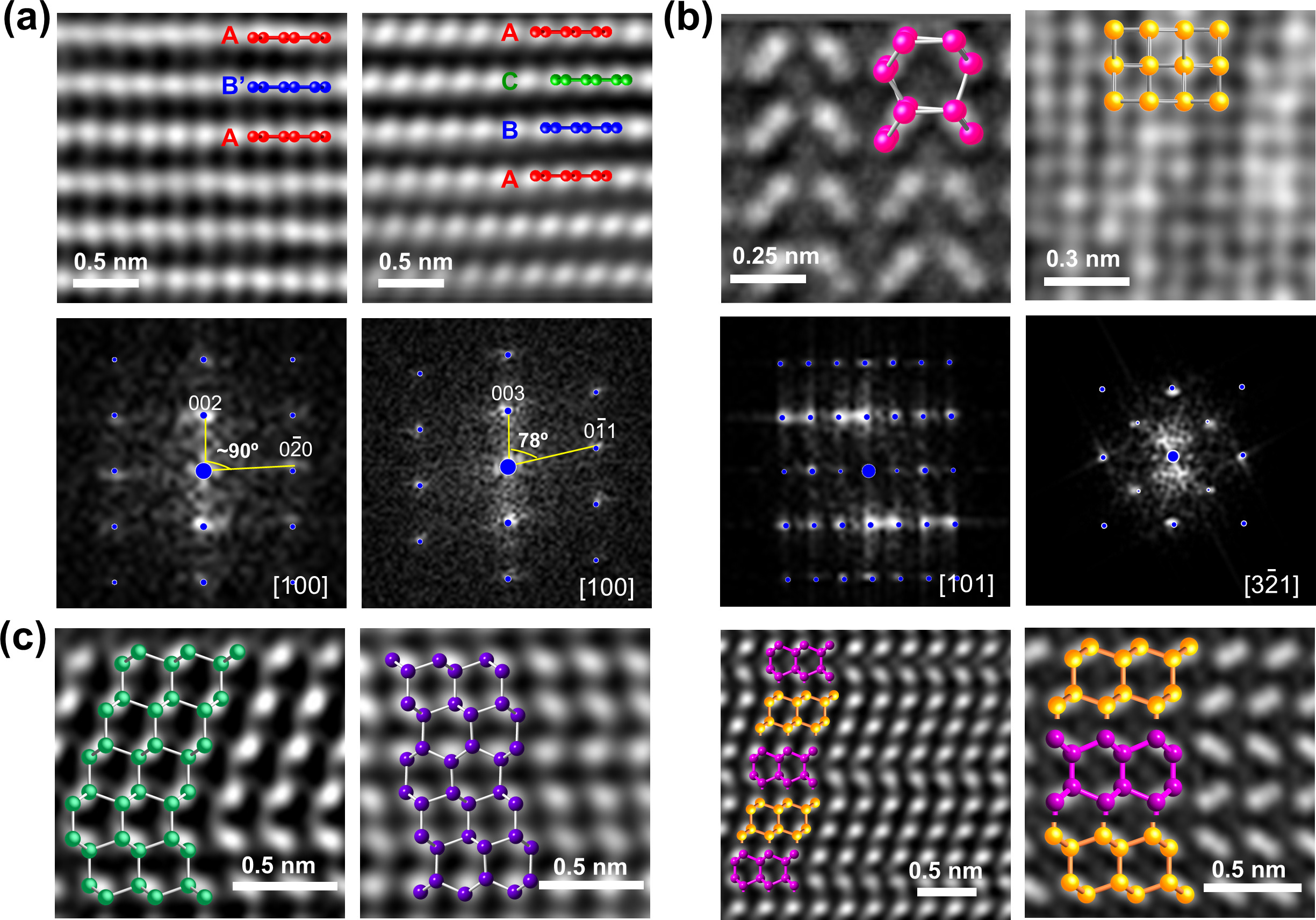}}
\caption{High-resolution TEM images of metastable phases of carbon along with the experimental and simulated diffraction patterns (blue circles).(a) Orthorhombic-graphtie with AB’ stacking and rhombohedral-graphite with ABC stacking; (b) Hexagonal-diaphite and cubic-Diaphite; (c) Intergrowth of hexagonal-diamond and cubic-diamond.}
\label{fig:TEM}
\end{figure*}

We use the information derived from the metastable phase diagram to explain the experimental observations during laser heating induced phase transformation of hexagonal graphite in a pressurized diamond anvil cell (DAC). We perform  HPHT experiment by loading a 60$\times$20 $\mu$m  single crystal graphite disk into a DAC. The pressure was first raised to 20 GPa monitored by ruby fluorescence. The graphite crystal was heated to $\approx$1400 K by a YAG laser at the center of crystal. Due to Gaussian distribution of laser spot, the temperature away from the center could be as low as $\approx$1000K, such that the center part of the sample was turned into dark transparent and the outside rim remains as dark. In this recovered sample, several metastable phases were identified by HRTEM as shown by the images in Figure \ref{fig:TEM}. When pressurized, the graphite layers slide with respect to each other to form orthorhombic and rhombohedral graphite (Figure \ref{fig:TEM}(a))~\cite{RN37,RN38,DuanHRTEM}. Around the dark rim near transparent areas, with further increase in temperature, the orthorhombic and rhombohedral graphite layers buckle to form interlayer bonds resulting in the formation of hexagonal or cubic diamond respectively~\cite{RN37,RN38,RN39,RN40,RN41,RN42,RN43,RN44,RN45}. In practice, both the transformation pathways occur simultaneously, resulting in an intergrowth of cubic and hexagonal diamond\cite{RN14,RN15,RN26,RN44}, also known as the stacking disorder (shown in Figure \ref{fig:TEM}(c)). 
The diaphite-like lonsdaelite phase 
with two different bond lengths (Figure \ref{fig:TEM}(b), supporting information section S8) was also observed after the HPHT treatment. One can explain this observation with the aid of our metastable phase diagram. The diaphite phase is easily accessible under the experimental conditions used (20 GPa, 1400 K) since it is metastable with only a $\Delta G$=45 meV/atom (Figure \ref{fig:phasediagram}(d); purple phase) which is $\approx k_B T/3$. We conjecture that graphite undergoes phase transformation, triggered by the excitation in experiments, into a accessible metastable phase which can be represented as excitation induced hopping from the global minima to a local minimum in the free energy landscape. 



Furthermore, we observe a new cubic-diamond like phase exhibiting the same diffraction pattern as the previously reported \textit{n}-diamond~\cite{RN1}. New diamond (\textit{n}-diamond) was proposed as a new carbon allotrope; its electron diffraction pattern matches that of cubic (Fd-3m) diamond apart from some additional reflections that are forbidden for diamond, indexed as \{200\}, \{222\} and \{420\}. The speculation of this new allotrope was first reported in 1991~\cite{RN1}, but the exact crystal structure of \textit{n}-diamond has remained as a controversy despite several attempts to explain the \textit{n}-diamond diffraction pattern\cite{RN46,RN47,RN48,RN49}. Here, we attempt to explain the crystal structure of the metastable \textit{n}-diamond using our metastable phase diagrams. Among all the phases that appear near the experimental conditions ($\approx$20 GPa, 1400K) in the metastable phase diagram at $\Delta G$ = 45 meV/atom (Figure \ref{fig:phasediagram}(e)), the diffraction pattern of S291 phase matches excellently with experiments (Figure \ref{fig:TEM}(b)). The S291 phase is a cubic analog of the diaphite-like lonsdaelite phase with two different bond lengths (section S6 in supporting information). Similar to the diaphite-like lonsdaelite phase, 
cubic-diaphite is dynamically stable and has no imaginary phonon modes under a highly anisotropic pressure (section S6 in supporting information). Such anisotropy in pressure is present in our experiments. In the dark area where graphite was not converted into diamond, we found graphite layers are severely bent and formed into many empty pockets with a rhombus shape. Under an anisotropic pressure, the atomic plane distance becomes much shorter at these bent areas, equivalent to a huge increase in pressure in the out-of-plane direction. It is predicted that diamond nucleates at these bent areas~\cite{Xie2014Graphite2Diamond}. 
 
Hence, our framework not only correctly reproduces the dominant diamond and graphite phase in the equilibrium phase diagram, but also explains the observation of metastable phases in HPHT experiments. Mapping the metastable phase diagram and inspecting the neighboring phases provides insight into possible phase transformations pathways and assists in selecting the appropriate starting material for targeted synthesis, thus accelerating computer-aided materials discovery.


\section{Discussion}

\subsection{Domains of relative stability}

\begin{figure*}
\noindent\makebox[\textwidth]{\includegraphics[width=\textwidth]{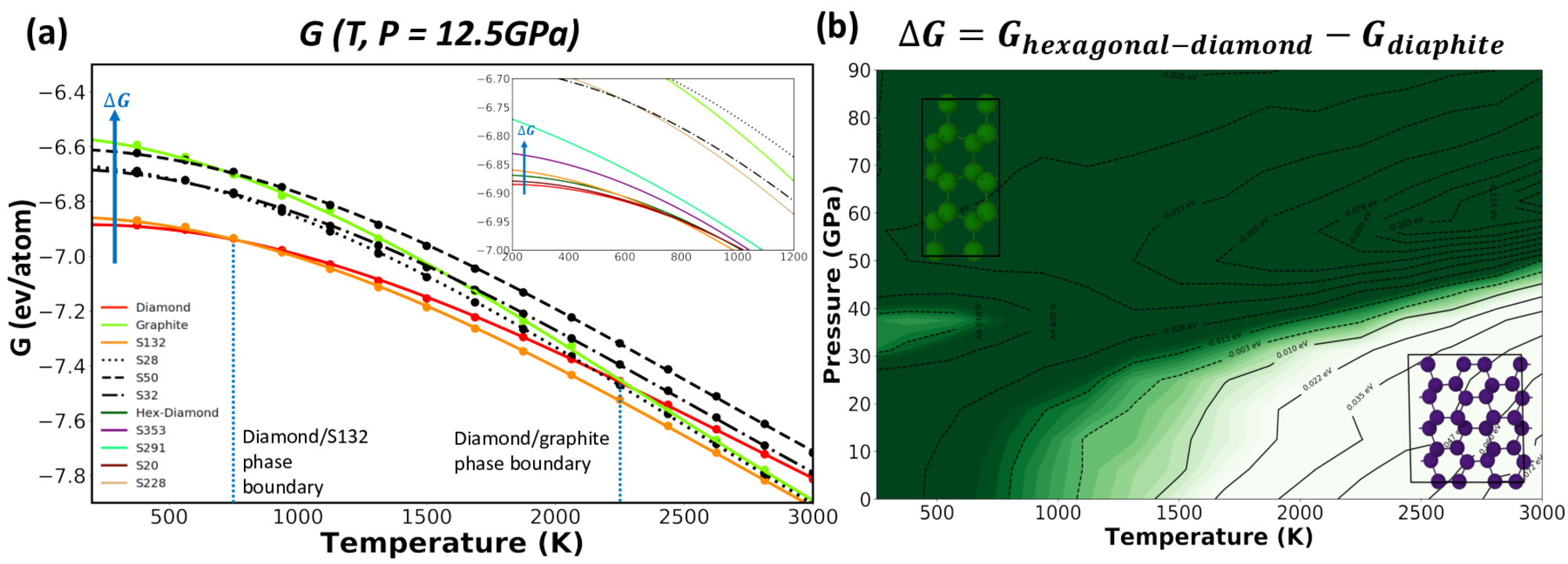}}
\caption{Relative stability and domains of metastability: (a) $G(T,P = 12.5 GPa)$  of equilibrium and some of the metastable phases; (b) Relative stability between hexagonal diamond (green) and diaphite (purple) computed as $\Delta G = G_{hexagonal}- G_{diaphite}$} 
\label{fig:domains}
\end{figure*}

The phase diagrams discussed above were generated by comparing the free energies of \textit{all} the candidate phases. Often, materials scientists find it useful to consider only a select few phases of interest and inspect their relative probability of formation. For example, one may consider only two phases involved in a phase transition and study their relative stability, to estimate the phase transition line. The probability of observing a phase at a given pressure and temperature depends on its relative stability with respect to the competing phases. Figure \ref{fig:domains}(a) shows the free energy profile, at $P$ = 12.5 GPa, of the phases that appear in equilibrium phase diagram and the near equilibrium metastable phases S20, S28, S32, S50, S132, S353, S291 and S228. The points where any two pair of lines intersect is the phase boundary between the corresponding phases. Free energies of distinct phases are separated by a finite $\Delta G$ (also the degree of non-equilibrium). The relative stability can also be considered as the projection, on the $T-P$ plane, of the distances between the free energy surfaces $G(T,P)$ for each phase. Figure \ref{fig:domains}(b) shows the map of the difference in the free energies $\Delta G=G_{hex-diamond}-G_{diaphite}$. Experimentally, diaphite is observed at moderate pressures and high temperatures whereas high pressure conditions predominantly yield hexagonal diamond. 
Such information about the relative stability can aid in driving the synthesis process to yield a desired metastable phase, as opposed to a mixture of phases, by appropriately tuning the experimental conditions.

\subsection{Domains of synthesizability}
\begin{figure*}
\noindent\makebox[\textwidth]{\includegraphics[width=\textwidth]{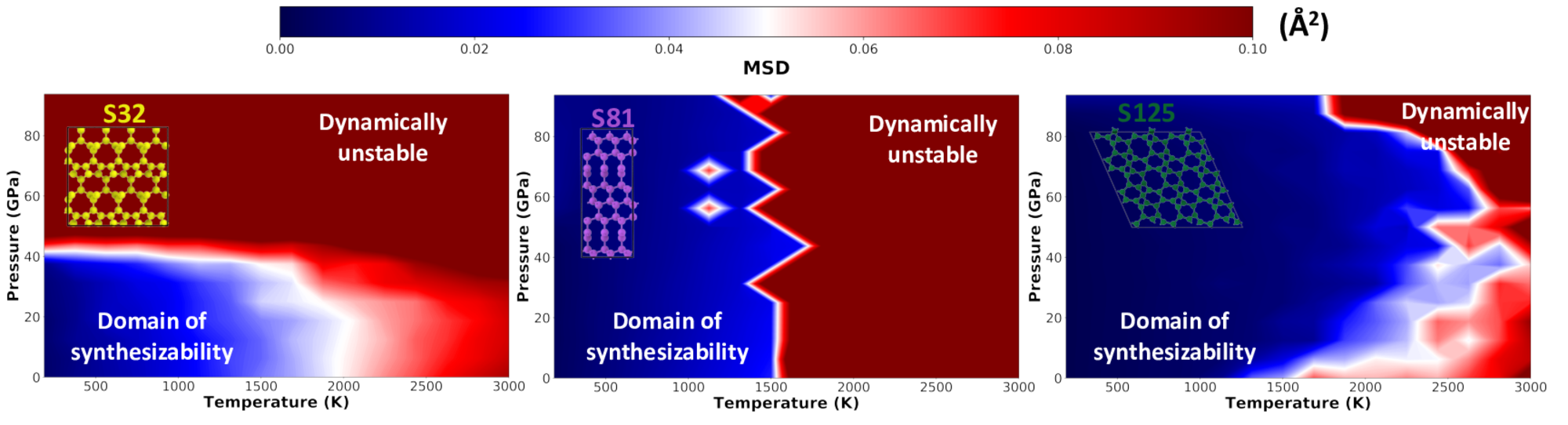}}
\caption{Domains of synthesizability based on dynamical stability of S32, S81 and S30 respectively}%
\label{fig:domainsSynthesizability}
\end{figure*}

The possibility of observing a phase at a given $T$ and $P$ depends on whether the crystal structure is retained or deformed due to melting or dynamical instability. In other words, the synthesizability is fundamentally limited by dynamical stability. We analyze the dynamical stability of the metastable phases using the mean square deviation (MSD) of the atoms during the MD simulations. A phase is dynamically unstable if the MSD is greater than 0.1 \AA. Here, we define domain of synthesizability as the region in the ($T,P$) space where a phase is dynamically stable. Figure \ref{fig:domainsSynthesizability} shows the domains of synthesizability of S32, S81 and S30. While the synthesizability of phase S32 and S125 is pressure limited, S81 is temperature limited. It should be noted that staying within the domain of synthesizability is a \textit{necessary}, but not a \textit{sufficient} condition for successful synthesis as there may be other factors limiting the synthesis. Similar upper limits for synthesizability, but based on the energetics of the amorphous phase, has been proposed in the past~\cite{KristinPearsonSynthesizabilityLimit}. When a metastable phase is driven into a region of dynamically instability, it may transform into a neighboring metastable phase in the energy landscape or undergo melting to form an amorphous phase. Such theoretical bounds on the state variables $(T,P)$, where a phase is likely to be stabilized, are instructive for the synthesis a metastable phase of interest.

\begin{figure*}[!htb]
\noindent\makebox[\textwidth]{\includegraphics[width=\textwidth]{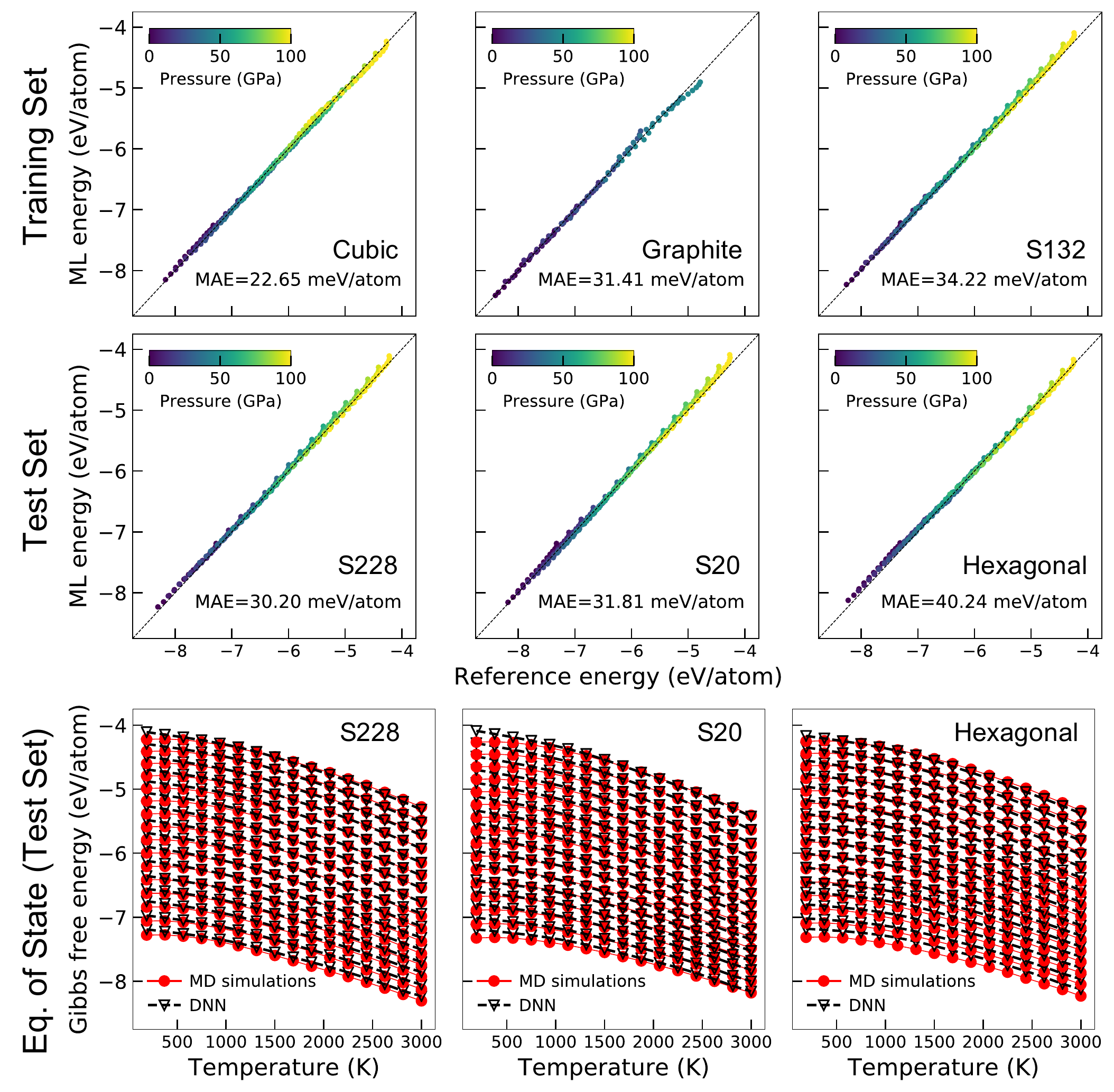}}
\caption{Performance of the DNN model to predict Gibbs free energy of different phases of Carbon. Parity plot demonstrating prediction accuracy of DNN model against reference Gibbs energy dataset for the  (a) training and (b) test sets. (c) Gibbs free energy predictions for the S228, S20 and hexagonal phases for various temperature and pressures. Although these phases were part of the test set, DNN predicts their energetics accurately.}
\label{fig:NN}
\end{figure*}

\subsection{Accelerating construction of metastable phase diagrams using machine learning}

The generation of metastable phase diagram relies on expensive free energy computations for a large number of competing phases. Using ML based surrogate models, we show that this process can be accelerated, and surrogate models that predict $G(T,P)$ can be constructed. Figure \ref{fig:NN} presents the performance of the DNN model trained to predict $G(T,P)$ given only the structural information in the form of MBTR descriptor. The parity plots in Figure \ref{fig:NN}(a) \& (b) demonstrate the prediction accuracy (mean absolute error, MAE) achieved by the DNN model on the training as well as the test set. Notably, hexagonal-diamond, S228 and S20 data were part of the test set and the good DNN performance for these cases illustrates its capability to capture the free energy surface of carbon. Further, in Figure \ref{fig:NN}(c) we show that our DNN model is able to accurately predict the equation of state of phases in the test set, given only their structural information. The overall root mean square error (RMSE) across all phases in the test set was 90 meV/atom (see section S3 in supporting information). In many cases, high errors in free energy predictions were observed at relatively higher pressures, as partially captured in Figure \ref{fig:NN}. Once such a surrogate model is trained, the free energy landscape of any new phase can be predicted orders of magnitude faster using only the structural information, thus, speeding up the process of constructing metastable phase diagrams. 


\section{Conclusion}

In summary, we report on an automated workflow that allows for construction of a “metastable phase diagram”. We introduce an alternate representation of metastable phases and their relative stability by providing a free energy scale which helps identify both the metastable phase location and its extent of non-equilibrium. Such a representation is far more informative with regard to designing experiments and accelerating the discovery of metastable phases, which often display exotic properties. Our workflow constructs the metastable phase diagram by combining several synergistic computational approaches including a structural search based on genetic algorithms, deep learning accelerated high-throughput free energy calculations and multiclass support vector machines to classify phase boundaries. We demonstrate the efficacy of our computational approach by using a representative single component carbon system, whose equilibrium and metastable phases have been well studied in the past. We successfully predict the equilibrium phase diagrams, and use the metastable phase diagram to explain several experimentally observed metastable intermediates including diaphitine-like lonsdaelite and its cubic analog, during high-pressure-high-temperature processing of graphite in diamond anvil cell. We also use the information extracted from the metastable phase diagram to propose a cubic-diaphitine structure, as a candidate phase to explain the diffraction pattern of \textit{n}-diamond. In addition, we show that the phase diagram construction can be accelerated by orders of magnitude with the help of a surrogate ML model, which can reliably predict the equation of states, given only the structural information. Our framework lay the groundwork for computer-aided discovery and design of synthesizable metastable materials. Our data-driven approach is fairly general and applicable to other chemical systems including multi-component alloys. For such systems, we envision a higher dimensional metastable phase diagram that allows exploration of metastability as a function of composition as well.


\section*{Acknowledgements}
Use of the Center for Nanoscale Materials, an Office of Science user facility, was supported by the U. S. Department of Energy, Office of Science, Office of Basic Energy Sciences, under Contract No. DE-AC02-06CH11357. This research also used resources of the Argonne Leadership Computing Facility at Argonne National Laboratory, which is supported by the Office of Science of the U.S. Department of Energy under contract DE-AC02-06CH11357. This research used resources of the National Energy Research Scientific Computing Center, a DOE Office of Science User Facility supported by the Office of Science of the U.S. Department of Energy under Contract No. DE-AC02-05CH11231. This material is based upon work supported by Laboratory Directed Research and Development (LDRD) funding from Argonne National Laboratory, provided by the Director, Office of Science, of the U.S. Department of Energy under Contract No. DE-AC02-06CH11357. We gratefully acknowledge the computing resources provided on Fusion and Blues, high performance computing clusters operated by the Laboratory Computing Resource Center (LCRC) at Argonne National Laboratory.

\ifarXiv
    

\section*{Supplementary material}

# Supporting Information
## Machine Learning the Metastable Phase Diagram of Materials

Srilok Srinivasan[1], Rohit Batra[1], Duan Luo[1], Troy Loeffler[1], Sukriti Manna[1,2], Henry Chan[1,2], Liuxiang Yang[3], Wenge Yang[3], Jianguo Wen[1], Pierre Darancet[1,4], and Subramanian Sankaranarayanan[1,2]

[1]Center for Nanoscale Materials, Argonne National Laboratory, Lemont, Illinois 60439, United States
[2]Department of Mechanical and Industrial Engineering, University of Illinois,Chicago, Illinois 60607, United States
[3]Center for High Pressure Science and Technology Advanced Research, Beijing 100193, P.R. China
[4]Northwestern Argonne Institute of Science and Engineering, Evanston, IL 60208, USA

November 29, 2021

# S 1 Detailed workflow for construction of phase diagrams

The detailed schematic our workflow is shown in Supplementary Figure 1. The inputs of our framework are the information about the chemical species, and the temperature and pressure range of interest. Given the chemical information and the range of $(T, P)$ we start our workflow by first identifying the metastable cystal phases using evolutionary structure search.

## S 1.1 Evolutionary structure search

Our evolutionary structure search is based on meta-heuristic genetic algorithm, wherein an initial gene pool of crystal structures are randomly guessed and evolved in the subsequent generations through genetic mutations or crossover between the fittest structures, which mimics Darwinian evolution.

We set the size of the gene pool as $N=40$. We initialize the gene pool with randomly guessed atomic positions and lattice parameters $(a, b, c, \alpha, \beta, \gamma)$, subject to the constrains:

1. no two atoms are closer than 0.5 Å

2. number of atoms in the unit cells lies between 4 and 20

3. length of the lattice vectors $(a, b, c)$ lie between 2 Å and 20Å

4. lattice angles $(\alpha, \beta, \gamma)$ lie between 20° and 160°

Structure search begins by computing the fitness of the initial gene pool structures based on their enthalpies after relaxing under a specified external pressure, with forces computed using density functional theory (DFT) using Perdew, Burke, Ernzerhof approximation.

In addition, we also perform independent evolutionary structure search using the long-range carbon bond-order potential (LCBOP)[7] model. Classical models like LCBOP are cheaper compared to DFT and allows for a quick search over the vast configurational space to identify the far-from-equilibrium metastable structures.

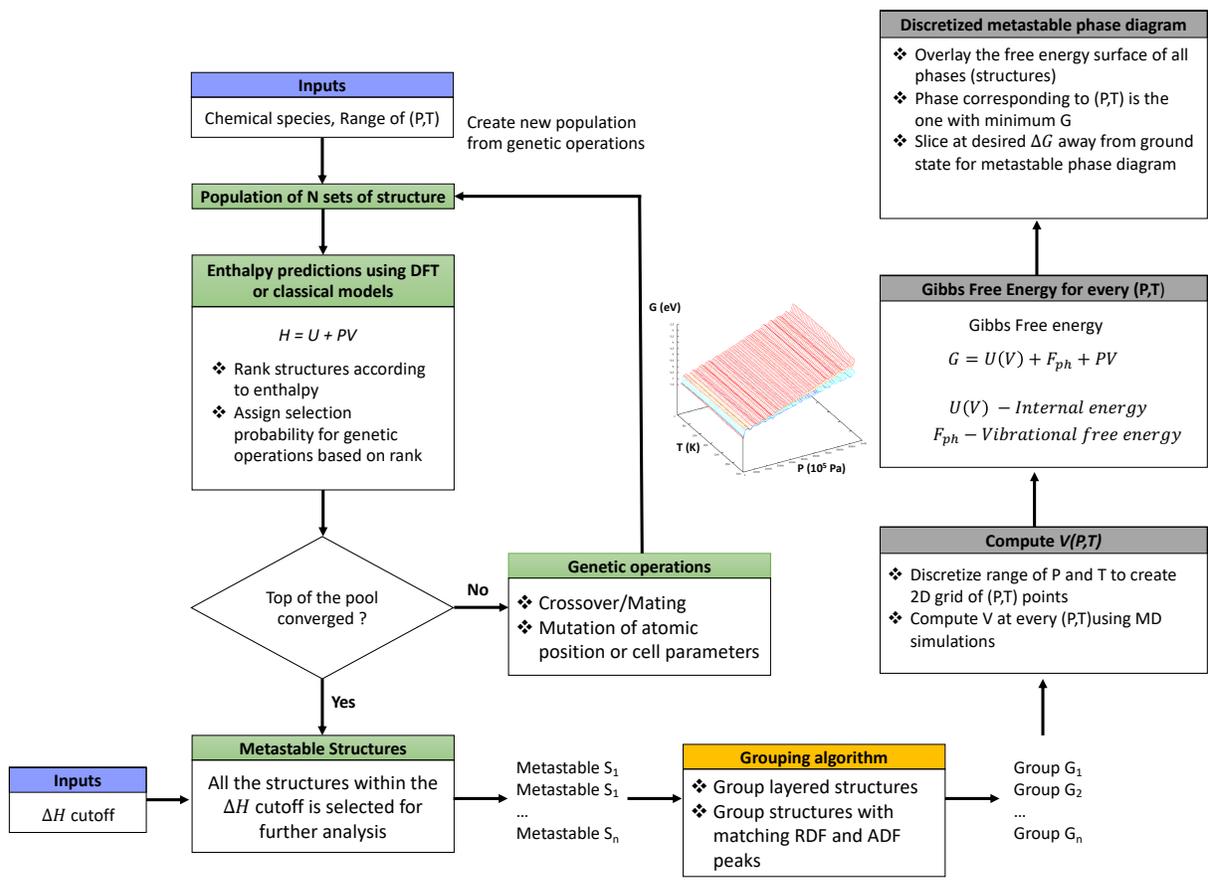

Supplementary Figure 1: Workflow for constructing metastable phase diagrams

The DFT relaxations are done using the VASP package [14]. LAMMPS package [19] is used to relax structures using LCBOP model. Fitness of each organism (structure) in a given gene pool is evaluated as

$$f_i = \frac{H_i - H_{max}}{H_{min} - H_{max}} \tag{S1}$$

where $H_i$ is the enthalpy of the organism $i$, $H_{max}$ and $H_{min}$ are the maximum and the minimum enthalpy in the current pool. The gene pool is ranked according to the fitness and parent structures are selected to undergo genetic variations to produce new offspring structures for the subsequent generation of gene pool. The selection probability of each structure is based on the fitness:

$$p_i = \frac{f_i}{\sum_i f_i} \tag{S2}$$

We define three types of genetic operations to build the subsequent generation of structures:

1. **Crossover variation**: This genetic variation involves two parents structures. The offspring structure is generated by slicing the parent structures across a random axis and combing the atoms on one side of the slice with the atoms on the other side in the other parent structure.

2. **Structure mutation**: Structure mutation involves random perturbation of the atomic coordinates and the lattice parameters

3. **Number of atoms mutation**: Atoms are randomly deleted or added in such a way that the constrains on inter-atomic distances and maximum number of atoms allowed.

A new generation offspring structures are generated using the above operations. The probability that a parent structure is subjected to crossover variation, structure mutation and number of atoms mutation are – 0.4, 0.4 and 0.2 respectively. Each of the offspring structure has to pass a redundancy check and satisfy the above mentioned constrains before it can be added to the gene pool of the next generation. Once a new gene pool of 40 structures are obtained, the fitness of the new generation is evaluated and new set of parents are selected based on their probabilities. This cycle is repeated until the difference between the enthalpy of the best and the top $N/8$ structure is less than a tolerance. The tolerance we used for the case of carbon is 20 meV. We build our algorithm based on the modules and function definitions within the *Genetic algorithm for structure and Phase prediction* code [2]. Further details on the algorithm and the genetic variations can be found in Ref. [20, 22].

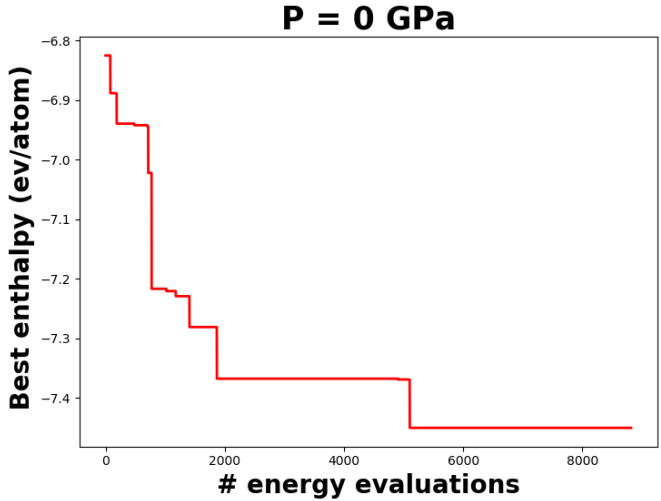

Supplementary Figure 2: Evolution of the best structure in the pool

We perform the independent evolutionary structure searches at $P = 0$ GPa, $P = 10$ GPa & $P = 100$ GPa. After convergence, we build a consolidated list of distinct structures ordered according to increasing value of enthalpies. Only the phases with satisfying $H < H_{ground} + \Delta H_{cut-off}$ are selected for free energy calculations. For carbon, the graphite phase is the experimental ground state with the minimum enthalpy of -7.365 eV/atom (computed using LCBOP model).

## S 1.2 Grouping based of RDF and ADF

Some of the candidate structures are structurally very similar and the enthalpies vary only by a small value. For example, in the case carbon, our structure search algorithm predicts hexagonal graphite, orthorhombic graphite and rhombohedral graphite, all which only differ in their stacking patterns and have very similar structural features. Besides at high pressure and temperature conditions, it is highly probable for the layers to slide against each other and change stacking, as can be seen in Figure 3 of the main text. Hence, we group such structures with very high similarity and count them as the same phase. i.e hexagonal graphite, orthorhombic graphite and rhombohedral graphite are considered as "graphite" phase. Only the candidate phase with the least enthalpy within each group is used to compute the free energies. The grouping is done based on the radial distribution function (RDF) and angular distribution function(ADF). Any two structures with matching first two peaks of RDF and ADF are grouped together. We end up with 505 unique groups within 670 meV from the ground state graphite phase. The free energies of the structures within the same group vary only by a small value (Supplementary Figure 3). The Crystallographic Information File (CIF) for each structure is provided in a GitHub repository.

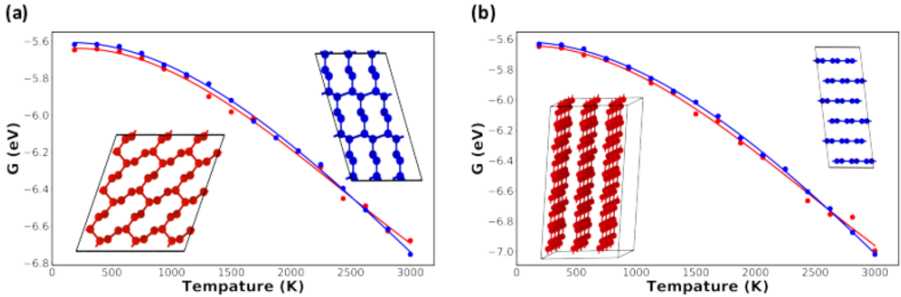

Supplementary Figure 3: Free energy profile of structures within the same group

## S 1.3 Free Energy Calculations

The temperature (0-3000 K) and pressure range (0-100 GPa), over which the phase information is desired, is discretized into a 16×16 uniformly spaced grid. The free energy of the candidate phases are computed at a temperature and pressure corresponding to each of the grid points. The Gibbs free energy can be written as

$$G(T_i, P_i) = H(T_i, P_i) - TS(T_i, P_i). \tag{S3}$$

where $T_i$ and $P_i$ are the temperature and pressure corresponding to a given grid point $i$

We first compute the enthalpy ($H$) and density($\rho$) of all phases at each of the grid point from MD simulations at the corresponding temperature and pressure ($T_i, Pi$). We thus perform 256 (16×16) MD simulations for each phase. We construct a super-cell of 3×3×3 of the initial unit cell to minimze the finite size effects. The interatomic interactions are modeled using the LCBOP [7] potential. We use a timestep of 1fs. The system is first equilibrated under an NVT ensemble with Nosé–Hoover thermostat at the target temperature of $T_i$ for 100ps before switching to an NPT ensemble, with target temperature ($Ti$) and pressure ($P_i$) controlled by Nosé–Hoover thermostat and barostat respectively. The system is subsequently equilibrated for 200ps, which is sufficient for the density (volume) to converge. We simulate the system for an additional 200ps to determine $H$ and $\rho$ by averaging over the data obtained every 10 timestep. All MD simulations were performed using the LAMMPS package[19].

The entropic contribution to the Gibbs free energy $(-TS)$ is determined by modeling the atomic vibrations as harmonic oscillators. The entropy of a system of harmonic oscillators can be written as:

$$-TS_{vibrational}(T_i, P_i) = F_{Harmonic} - U_{Harmonic} = k_B T \sum_{q_i v_i} \ln[1 - \exp(-\frac{\hbar \omega(q_i v_i)}{k_B T_i})] \qquad (S4)$$

At each grid point, the phonon spectrum computed at the corresponding equilibrium density $(\rho_i)$ obtained from the MD simulations. The phonon calculations were performed using the PHONOPY package [23]. The force matrix for the phonon calculations are obtained from the LCBOP model. The total Gibbs free energy $G(T_i, P_i)$ is obtained by summing $H(T_i, P_i)$ and $-T_i S_i$.

## S 2 Stability of stacking disorder (S132) and hex-diaphite (S353)

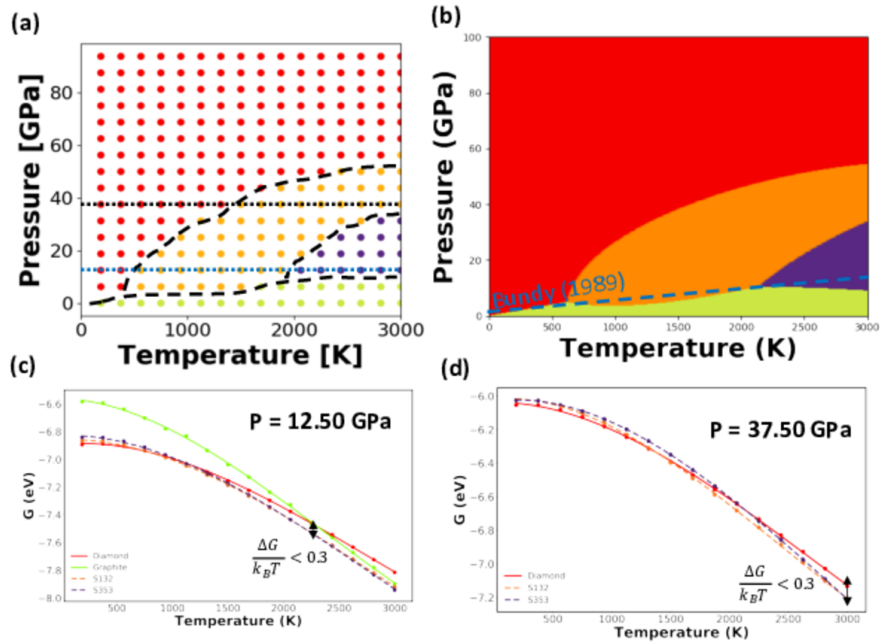

Supplementary Figure 4: Equilibrium phase diagram including the stability of S132 & S353

Supplementary Figure 4 (a) & (b) shows the discretized and continuous equilibrium phase diagram constructed by comparing the $G(T,P)$ of *all* the candidate phases identified by our algorithm. Apart from cubic diamond and graphite, which are the dominant stable phases in the experimental phase diagram, we note the appearance of S132 and S353 near the phase boundary. Our calculations show that S132 and S353 has a range of marginal stability $(\Delta G/k_B T < 0.3)$ near the graphite-diamond phase boundary. Observation of metastable diamond and metastable graphite near the phase boundary has been reported in the past[3, 25, 8, 12, 18, 15, 26, 6, 21, 4, 5, 16, 17]. S132 is a stacking disorder phase consisting of alternating layers of cubic diamond and hexagonal diamond (orange in Figure 2 in main text). S353 is diaphitine like distorted hexagonal diamond (purple in Figure 2 in main text) consisting of atomic configuration with two different bond lengths at 1.47 Å and 1.53 Å Both the phases have been observed experimentally during high pressure high temperature treatment of graphite in a diamond anvil. We further inspect the $\Delta G$ between S132, S353 and the stable phases. Supplementary Figure 4(b)&(c) shows the Gibbs free energy $G(T, P = 12.50 GPa)$ and $G(T, P = 37.50 GPa)$ across the blue and black dashed lines in Supplementary Figure 4(a). The maximum difference with respect to the stable phase are at 3000 K with $\Delta G^{S132}_{Graphite}(T = 3000K, P = 12.5GPa) = -63$ meV/atom, $\Delta G^{S353}_{Graphite}(T = 3000K, P = 12.75GPa) = -66$ meV/atom, $\Delta G^{S132}_{Diamond}(T = 3000K, P = 37.50GPa) = -49$ meV/atom and $\Delta G^{S353}_{Diamond}(T = 3000K, P = 37.50GPa) =$

5

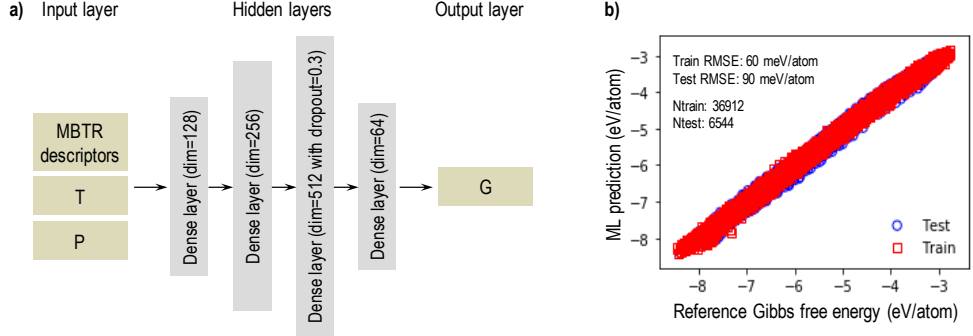

Supplementary Figure 5: a) Architecture of the DNN model used to learn Gibbs free energy and b) parity plot between the reference Gibbs free energy and respective DNN predictions for various phases of carbon in the training and the test set. The root mean square error (RMSE) in the DNN predictions for the training and test set are also included.

−12 meV/atom. At 3000 K, $\Delta G/k_B T < 0.3$ suggesting a very high probability of forming these phases along with diamond and graphite at those conditions. Our results are in agreement with the experimental conditions under which stacking disorder phase and diaphite phase are experimentally observed to co-exist with diamond and graphite [3, 25, 8, 12, 18, 15, 26, 6, 21, 4, 5, 16, 17].

## S 3    Deep Neural Network

A deep neural network (DNN) was used to learn the Gibbs free energy of different phases of carbon. It consisted of 4 fully connected (dense) hidden layers with 128, 256, 512 and 64 neurons, respectively, as shown in Supplementary Figure 5(a). The input layer consisted of many-body tensor representation [11] (MBTR) of the 0 K and 0 GPa structure of a phase, and the normalized T and P value. The MBTR fingerprint was obtained using the python library Dscribe [9] with 25 dimensions for the 'k2' (min=0.1, max=2 and $\sigma$=0.1) and 25 dimensions for the 'k3' (min=0, max=180 and $\sigma$=5) type terms, each normalized individually to the Euclidean length ($L^2$). Features with zero variance throughout the data were removed, while T and P values were included as two additional features, overall resulting in a 43-dimensional input fingerprint to the DNN.

The output layer consisted of a single neuron describing the DNN predicted Gibbs free energy of a phase at the input T and P values. The DNN was trained using Adam optimization algorithm [13] with the mean absolute error chosen as the loss function definition. Free energy data corresponding to 248 phases was used to train the model, while that for 30 and 43 phases was used as the validation and test set, respectively. A few important phases, such as cubic, graphite, S132, S291 and S353, were part of the training set, while others, including hexagonal, S228 and S20, were part of the test set. Since some phases were found to be dynamically unstable at different P and T conditions, caution was taken to only include those data points that correspond to reasonable free energy values without any arbitrary jumps in the free energy vs T, or free energy vs P behavior. The number of training epochs was determined by monitoring the model performance on the validation set and a dropout layer (with value of 0.3) was used after the third hidden layer for regularization purposes. The DNN code was implemented in Tensorflow [1]. The overall performance of the DNN model on the training as well as the test set is presented in Supplementary Figure 5(b).

In particular, using our surrogate ML model, we can quickly estimate the proximity of a newfound metastable phase with respect to the ground state, given only the structural information. The probability of realizing a metastable phase at a given temperature and pressure is directly proportional to $\exp(-\frac{\Delta G_{GS_i}^{MS_j}}{k_B T})$ with $\Delta G_{GS_i}^{MS_j} = G_{MS_j} - G_{GS_i}$ where $GS_i$ and $MS_j$ are the ground state and the metastable phase of interest. Supplementary Figure 6 shows the predictions of the DNN at 12.5 GPa, $G(T, P = 12.5 GPa)$, for metastable phases the ML model has never seen during training (S228, S20 and S50 are part of test set). The error

6

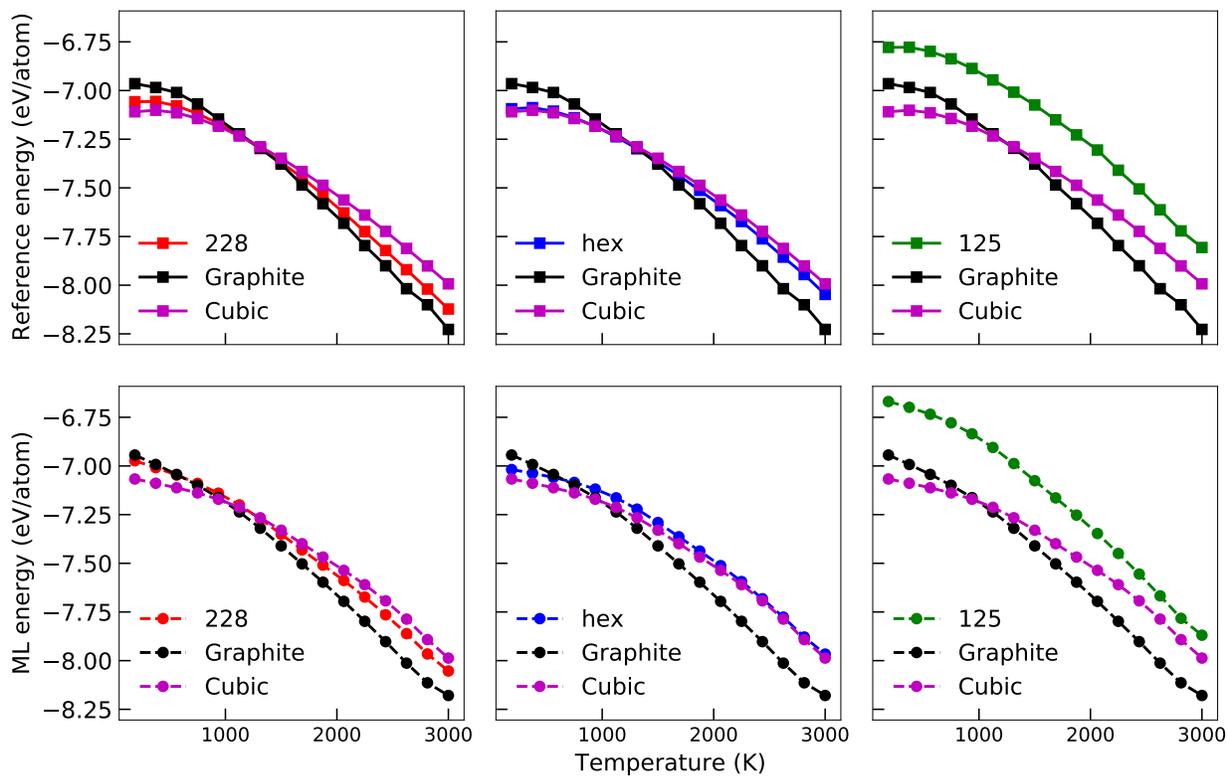

Supplementary Figure 6: (a),(b),(c) $G(T, P = 12.5 GPa)$ computed using our workflow for S228, S20 & S125 respectively. Cubic diamond and graphite are plotted alongside for comparison (d),(e),(f) $G(T, P = 12.5 GPa)$ computed using DNN for S228, S20 & S125 respectively. Cubic diamond and graphite are plotted alongside for comparison. S228 and S20 are near equilibrium and S125 is far from equilibrium

7

between MD computations and the ML predictions are less than 40 meV/atom. Thus, we can quickly classify a metastable phase as near-equilibrium and more likely to be synthesized, or far-from-equilibrium and less likely to be synthesized, by comparing free energies with ground state phases.

# S 4 Far from equilibrium metastable phase diagrams

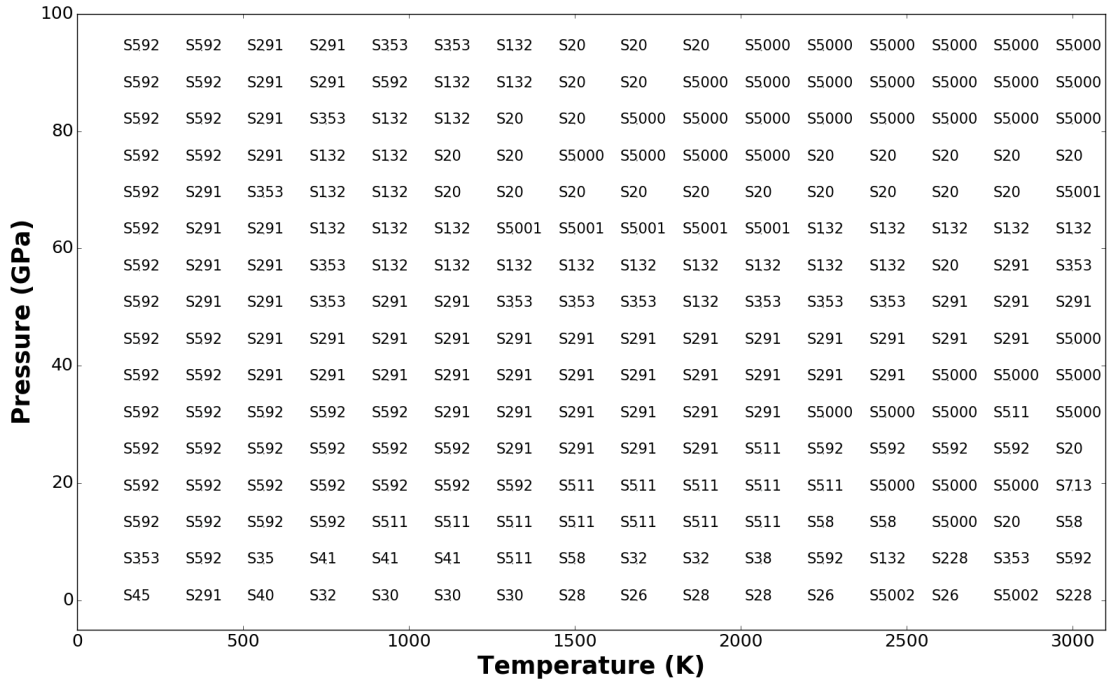

(a) $\Delta G = 100 meV/atom$

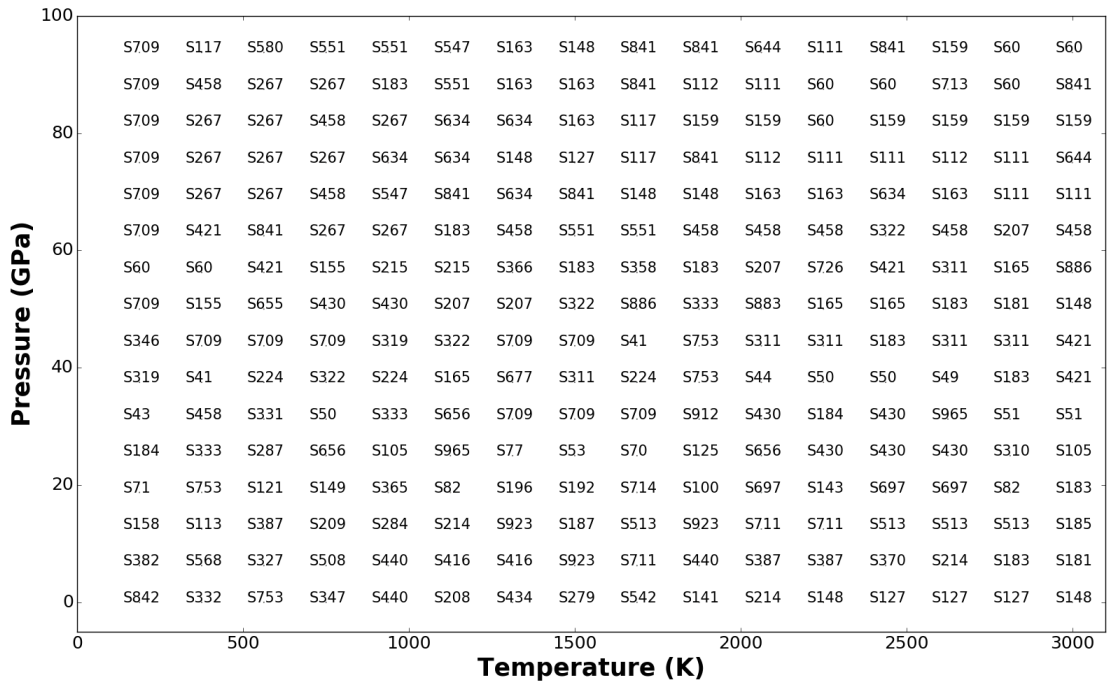

(b) $\Delta G = 500 meV/atom$

Supplementary Figure 7: Far-from-equilibrium metastable phase diagram

9

# S 5  High pressure high temperature processing

HPHT samples were obtained using a diamond anvil cell (DAC). Starting material is a 60×20 $\mu$m single crystal graphite disk cut from a millimeter size crystal by micro laser drilling system and it was loaded into a hundred micron diameter rhenium gasket chamber. Pressure was monitored by ruby fluorescence. When pressure is at 20 GPa YAG, laser heated samples at the center ( 1400 K) turned dark transparent but the rim remained dark. In this work, after decompression from high pressure and temperature treatment, we opened the DACs, transferred the samples from the chamber to a clean marble mortar with a tiny pin. TEM samples were prepared by crushing the recovered sample using a marble mortar and pestle and then dispersing these crushed powders onto a holey carbon grid. Focused-ion beam (FIB) technique is also used to prepare plane-view and cross-sectional TEM specimens. Argonne Chromatic Aberration-corrected TEM (ACAT, FEI Titan 80-300ST TEM/STEM) with a field-emission gun was used to investigate the crystallographic orientation, high-resolution transmission electron microscopy (HRTEM) images from the recovered samples.

# S 6  $n$-diamond

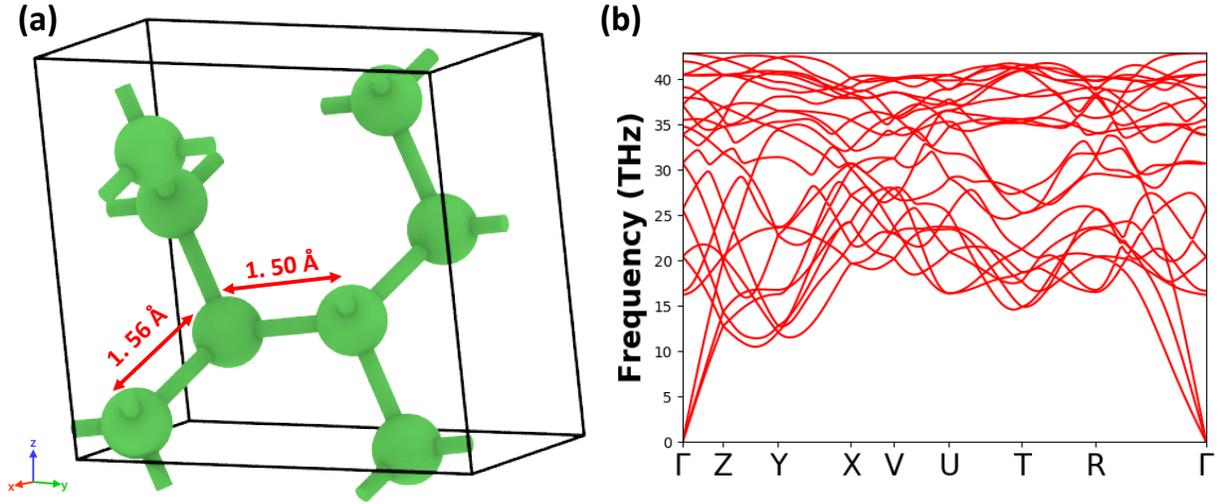

Supplementary Figure 8: (a): Structure of $n$-diamond after relaxation, (b) phonon dispersion of $n$-diamond

The initial structure of $n$-diamond (S291) as identified by our evolutionary algorithm is relaxed under an anisotropic pressure of 48 GPa in the $y$-direction and 20 GPa in the $x$- and $z$- directions. The resulting structure is still a cubic diamond like structure with two different bond lengths of 1.56 Å and 1.50 Å(Supplementary Figure 8). The simulated diffraction pattern of the final structure matches well with the previously reported $n$-diamond structure. We next inspect the stability of the proposed structure by computing the phonon spectrum and checking for any possible imaginary modes. The phonon spectrum is computed using PHONOPY package [23] with force constants obtained form density functional perturbation theory(DFPT). The relevant high symmetry points labeled in the phonon spectrum were obtained using the algorithm described in Ref.[10] which uses the spglib library [24] to construct the brillouin zone. The structure is stable since there are no imaginary modes.

Cubic-diamond consists of two fcc lattices that shift along [111] diagonal direction respect to each other. When the shift distance equals to a $sp^3$ bond length, (200) diffraction spots extinguish since cubic-diamond has one $sp^3$ bond length. When the shift distance of these two fcc lattices is away from 1.54 Å, the intensity at (200) diffraction spots gradually increases. In the simulated diffraction pattern using S291 structure, we can find the intensity of (200) spots are much lower than (400) spots due the small difference in these two bond lengths (1.56 Å, 1.50 Å). In the experiment diffraction pattern, the (200) intensity is close to the (400), indicating this n-diamond has a much smaller bond length than 1.50A (close to real graphite $sp^2$ bond length 1.42 Å).

# S 7 Relative stability of far from equilibrium structures

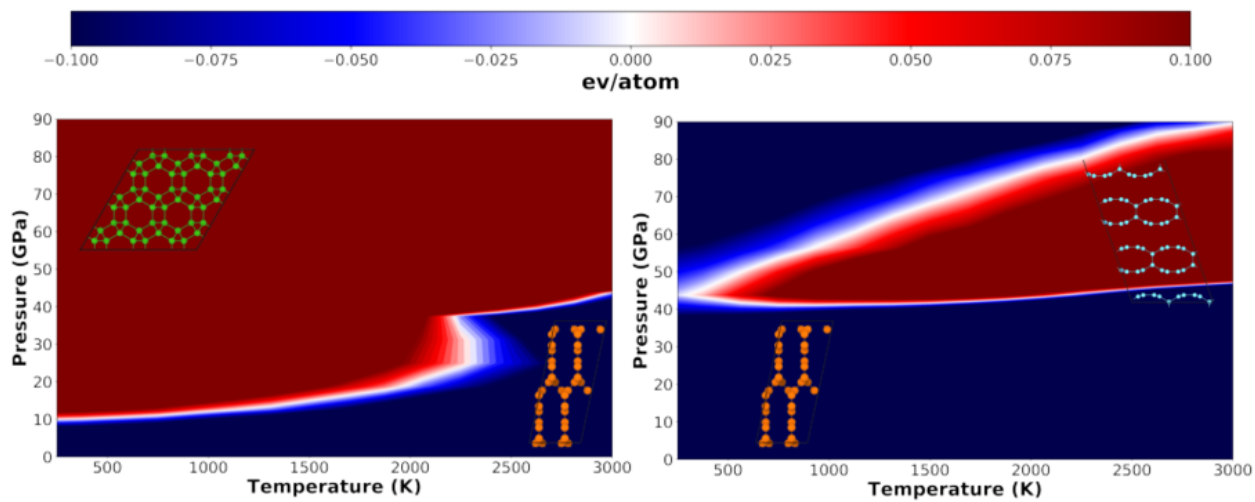

Supplementary Figure 9: Relative stability of far from equilibrium structures

# S 8    Diaphite-like lonsdaelite phase

Evidence of diaphite-like lonsdaelite phase are provided in Ref [27]. Specifically, Figure 2 of Ref [27] shows the AC-HRTEM image of lonsdaleite along [11$\bar{2}$0] with two different bond lengths ($OA \approx 1.56$ Å and $OB \approx 1.47$ Å) and Lonsdaelite phase after relaxing under anisotropic pressure. The UV Raman spectrum on the recovered sample shows three peaks (Figure 3 of Ref [27]). The DFT calculated frequencies dependence on bond lengths match with that of experiments.


\fi

\end{document}